\documentclass[fleqn,usenatbib]{mnras}

\usepackage{times}

\usepackage[T1]{fontenc}

\usepackage{graphicx}	
\usepackage{lipsum}
\usepackage{pdflscape}
\usepackage{afterpage}
\usepackage{multirow}
\usepackage{makecell}

\newcommand{\src}{GX~339--4}
\newcommand{\hxmt}{\textit{Insight}--HXMT}

\newcommand{\nustar}{\textit{NuSTAR}}
\newcommand{\swift}{\textit{Swift}}

\title[GX~339-4]{Rapidly alternating flux states of GX~339--4 during its 2021 outburst captured by \textit{Insight}--HXMT}

\author[Liu et al.]{Honghui Liu,$^{1}$
Jiachen Jiang,$^{2}$ 
Zuobin Zhang,$^{1}$ 
Cosimo Bambi,$^{1}$\thanks{E-mail: bambi@fudan.edu.cn}  
Long Ji,$^{3}$
\newauthor
Lingda Kong,$^{4,5}$ 
and Shu Zhang$^{4,5}$
\\
$^{1}$Department of Physics, Fudan University, 200438 Shanghai, China\\
$^{2}$Institute of Astronomy, University of Cambridge, Madingley Road, Cambridge CB3 0HA, UK\\
$^{3}$School of Physics and Astronomy, Sun Yat-Sen University, 519082 Zhuhai, China\\
$^{4}$Key Laboratory for Particle Astrophysics, Institute of High Energy Physics, Chinese Academy of Sciences, 100049 
Beijing, China\\
$^{5}$University of Chinese Academy of Sciences, Chinese Academy of Sciences, 100049 Beijing, China}

\date{Accepted XXX. Received YYY; in original form ZZZ}

\pubyear{2021}

\begin{document}
\label{firstpage}
\pagerange{\pageref{firstpage}--\pageref{lastpage}}
\maketitle

\begin{abstract}
The low mass X-ray binary GX~339--4 entered a new outburst in 2021. At the end of the hard to soft transition of this outburst, \hxmt\ found that the source rapidly alternated between low flux and high flux states on a timescale of hours. Two high flux states lasted only for a period comparable to the orbital period of the observatory. Time-resolved spectral analysis shows that the sudden changes of flux are confined in the hard X-ray band (>4~keV). The variable non-thermal emission, including the power-law continuum from the corona and the reflected emission from the inner accretion disk, is responsible for the observed variability. The strength of the disk thermal emission and the inner radius of the accretion disk are consistent between the two flux states. Assuming the lamppost geometry, our best-fit disk reflection models suggest a very low corona height (within 3~$R_{\rm g}$) and there is no evidence of significant variation in the corona geometry either. The observed rapidly alternating flux states suggest that the intrinsic power of the corona must change during the state transition. We discuss possible mechanisms for the observed sudden changes in the coronal power of \src. 
\end{abstract}




\section{Introduction}

Low mass X-ray binaries (LMXRBs) consist of a compact object (black hole or neutron star) and a low-mass companion star. The systems are powered by accretion of the compact object through Roche lobe overflow and release most of the energy in X-rays. Black hole LMXRBs often spend most of their time in a quiescent state and sometimes go into an outburst that can last a few months. The length of the quiescent phase differs from source to source. During an outburst, the X-ray luminosity of the source can increase by several orders of magnitude \citep[e.g.][]{Remillard2006}, making them bright X-ray sources in the sky.

The outburst of a black hole X-ray binary (BHXRB) can be traced by plotting its X-ray hardness (e.g., count rate in hard X-ray band divided by that of soft X-ray band) and intensity (e.g., count rate of the detector) on the hardness intensity diagram (HID) \citep[e.g.][]{Belloni2005}. The relative contributions of thermal and non-thermal emission components change along with the outburst and produce distinctive spectral states \citep[e.g.][]{Remillard2006}. At the early stage of an outburst, the source is in a low hard state (LH) located in the lower right corner of the HID and the X-ray spectrum is dominated by non-thermal emission in the shape of a hard power-law. A steady jet is often observed in this state \citep{Fender2004}. The spectrum remains hard as the source gets brighter and, at some point, moves into a thermal emission dominated soft state in a few days. During this transition, the contribution of the thermal emission to the spectrum gradually gets stronger and the source goes through the hard intermediate state (HIMS) and soft intermediate state (SIMS). At the brightest phase of the outburst, some sources show a very high state (or steep power-law state) \citep[e.g.][]{Done2007} where the disk emission is associated with a strong and steep power-law component. After spending some time on the soft state, the source returns to the LH state following a reverse order.

GX~339-4 is a typical LMXRB discovered in 1973~\citep{Markert1973}. It goes into bright outburst every a few years and has thus been extensively studied at all wavelengths. \cite{Heida2017} studied its near-infrared spectrum and constrained the mass of the black hole to 2.3--9.5~$M_{\odot}$. This range is consistent with the measurement from spectral and timing analysis of its X-ray data ($9.0_{-1.2}^{+1.6}~M_{\odot} $\citep{Parker2016}; 8.28--11.89~$M_{\odot}$ \citep{Sreehari2019}). The distance to \src{} is still uncertain and \cite{Heida2017} only derive a lower limit of 5~kpc. A recent work by \cite{Zdziarski2019} suggests a distance of 8--12~kpc and a black hole mass of 4--11~$M_{\odot}$. The distorted iron K$\alpha$ line allows to determine the black hole spin parameter through its broad X-ray spectrum and it has been confirmed the black hole has a high spin \citep[e.g., $a_* \sim 0.95$][]{Parker2016}.

\begin{figure}
    \centering
    \includegraphics[width=\linewidth]{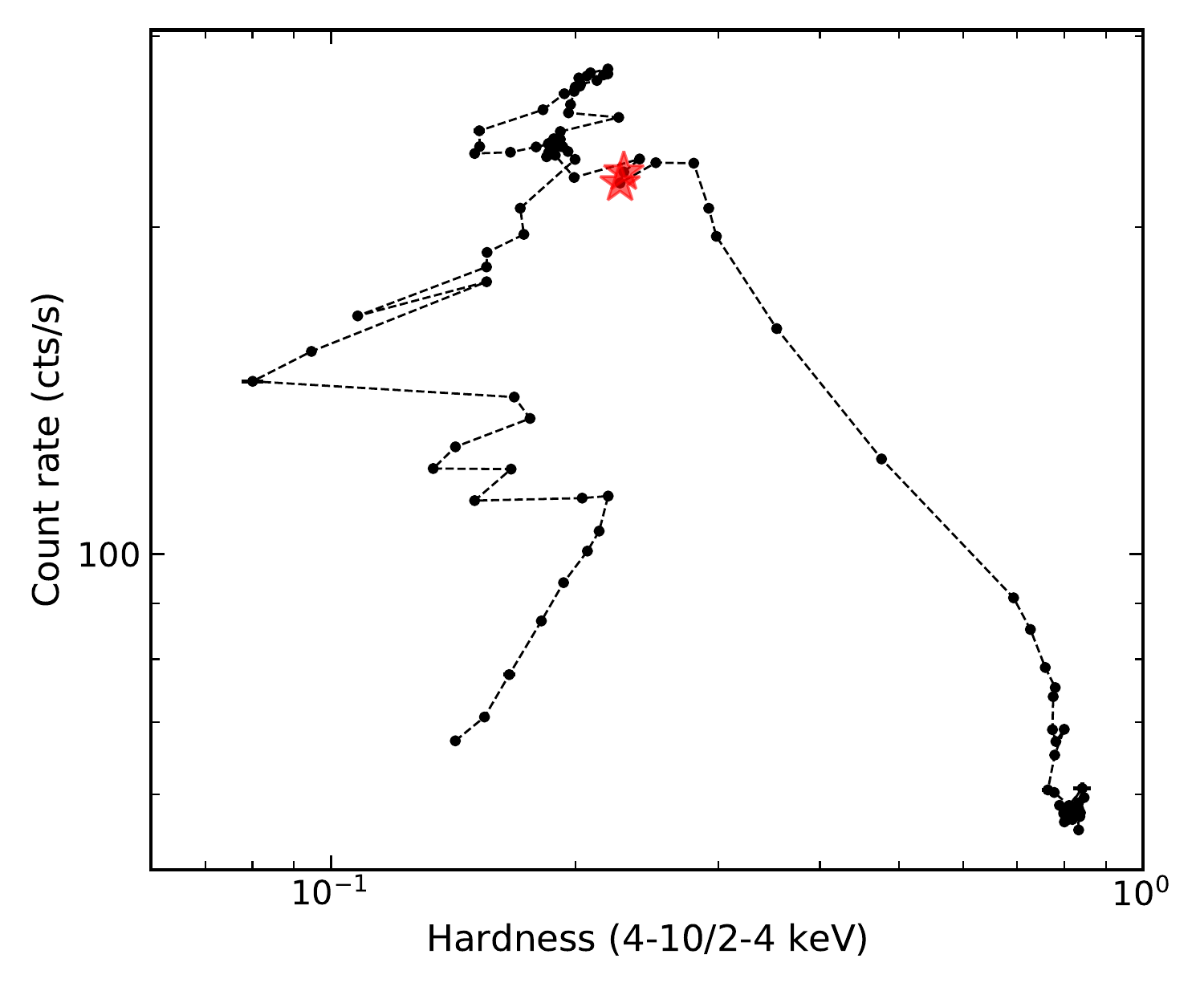}
    \caption{The hardness intensity diagram of the 2021 outburst of \src{}. Data are extracted from \hxmt{} observations. Each point represents data from one day. The two observations analyzed in this work are marked with red stars. The hardness is defined using the count rate.}
    \label{hid}
\end{figure}

\src{} entered a new outburst in 2021, which was captured by \hxmt{}. Fig.~\ref{hid} shows the HID of this outburst. The source spent some time in the hard state followed by a fast transition (around a week) to the soft state. At the end of this transition, where we see the red stars, we find that this source shows rapid changes of flux in the hard X-ray band as shown in Fig.~\ref{lcurve}. This variability happens on a timescale of less than one orbit of \hxmt{} (1.5 hours). After that, the source entered the very high state and then the soft state. The transition from soft to hard state was not observable by \hxmt{}.

In this paper, we investigate the hard X-ray alternating states by analyzing its time-resolved broadband spectra from \hxmt{}. We present the data reduction procedure in Sec.~\ref{reduction}. Overview of the alternating X-ray flux states is presented in Sec.~\ref{states}. The spectral fitting results are described in Sec.~\ref{analysis}. We discuss the results in Sec.~\ref{discussion}.

\section{Observations and data reduction}
\label{reduction}

\begin{table}
    \centering
    \renewcommand\arraystretch{1.5}
    \caption{\hxmt{} observations of \src{} analyzed in this paper.}
    \label{info-obs}
    \begin{tabular}{cccc}
        \hline\hline
        Date & obsID & \multicolumn{2}{c}{Exposure (ks)$^1$} \\
         & & LE & ME \\
        \hline
        2021-04-01 & P0304024043\textbf{01-07} & 5.4 & 15.0   \\ 
        \hline
        2021-04-02 & P0304024043\textbf{08-15} & 5.9 & 15.0  \\
        \hline

    \end{tabular}\\

\textit{Note}. (1) We show the total exposure time of all observations in one day.
\end{table}

\textit{Insight}-HXMT is the first Chinese X-ray telescope. it consists of low-energy (LE), medium-energy (ME) and high-energy (HE) detectors that cover the energy range of 1-250 keV \citep{Chen2020, Cao2020, Liu2020, Zhang2020}.  The light curves and spectra are extracted following the official user guide\footnote{\url{http://www.hxmt.cn/SoftDoc/67.jhtml}} and using version 2.04 of the software {\sc HXMTDAS}. We estimate the background using the scripts \texttt{hebkgmap}, \texttt{mebkgmap} and \texttt{lebkgmap} \citep{Liao2020a, Guo2020, Liao2020b}. Good time intervals are screened with the recommended criteria, i.e., the elevation angle $>$ 10 degrees, the geomagnetic cutoff rigidity $>$ 8 GeV, the pointing offset angle $<$ 0.1 and at least 300 s away from the South Atlantic Anomaly (SAA).

The LE data in the 2--9 keV band and ME data in the 8--15 keV band are used in this work. Data above 15~keV are ignored because of the very high background. For the use of $\chi^2$ statistics, spectra are binned to ensure a minimal counts of 30 for every energy bin.

\section{Alternating X-ray Flux States}
\label{states}

\begin{figure*}
    \centering
    \includegraphics[width=\linewidth]{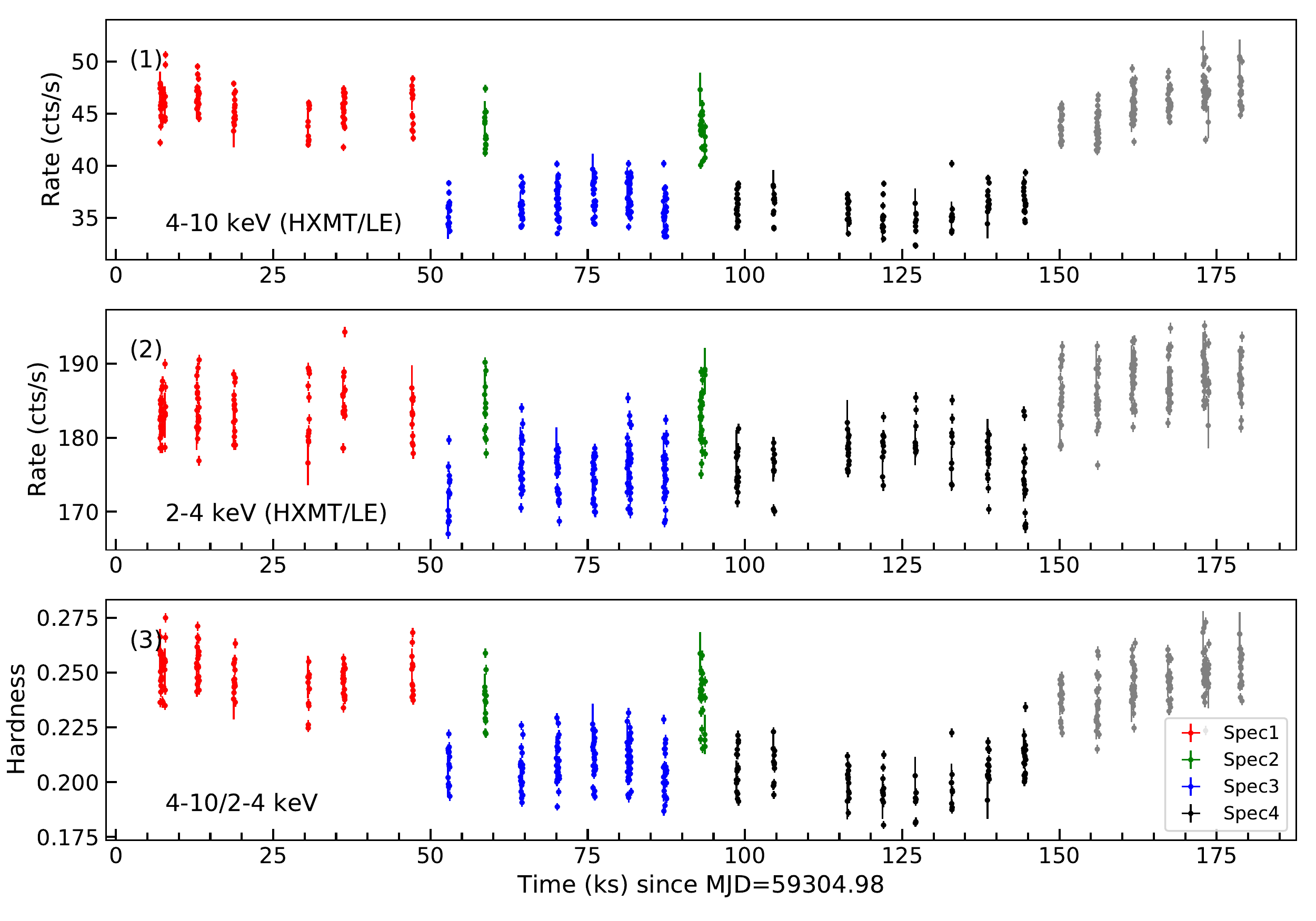}
    \caption{Light curves of \src{} by \hxmt{} in 4--10 keV (panel 1) and 2--4 keV (panel 2). The evolution of the hardness ratio is shown in panel 3. The source enters a low flux state at 52~ks that lasts for about 100~ks. Two brief high flux states emerge from the low flux interval on a timescale less than one orbit. The flux change is more prominent in 4--10 keV. Different colors mark the four spectra extracted from the observation (see Sec.~\ref{analysis}).}
    \label{lcurve}
\end{figure*}

In Fig.~\ref{lcurve}, we show the light curves of \src{} on the 1st and 2nd April 2021. Their corresponding hardness and intensity on the HID are marked in Fig.~\ref{hid}. The gaps in the light curves are due to the low Earth orbit of the telescope or the South Atlantic Anomaly (SAA). 

Panel (1) shows that the observed 4--10\,keV count rate of \src{} had a sudden decrease by a factor of 1.28 at 52~ks since the beginning of the observations. Then \src{} returned to the original high flux level during the next orbit of observation at 58~ks. A similar second brief high flux state happened around 93~ks. The first and second high flux states, which are marked in yellow in Fig.~\ref{lcurve}, showed similar observed X-ray flux. 

The 4--10\,keV count rate changed by a factor of 1.28, suggesting that the hard X-ray flux changed more significantly than the soft X-ray flux. The X-ray hardness of \src{} defined as the count rate ratio between the 4--10\,keV and 2--4\,keV bands was consistently higher in the high flux state than that in the low flux state.

The two brief high flux states lasted for a period shorter than the orbital period of the telescope. The transition between the high and low flux states had to be very rapid, e.g., less than the orbital period. Therefore, \hxmt{} did not observe the transition between the high and low flux states\footnote{The \swift\ observation of \src\ captured the state transition from the high to the low flux states. See Appendix \ref{otherlc} for more details.}. 

To ensure the observed high flux states were not caused by unknown instrumental issues, we checked observations from other missions in the archive. In particular, the second rise of flux was captured by other missions, including \nustar{} and \swift{} (see Fig~\ref{otherlc}). In this work, we focus on the \hxmt{} observations of \src{} only. We refer interested readers to detailed spectral analysis of \nustar{} and \swift{} data in the future work by Garcia et al.

In summary, we find that \src{} rapidly alternated between two flux states in the X-ray band during the intermediate state. The high flux states lasted for a short period that is comparable to the orbital period of the telescope. During the observed alternating flux states, the 4--10\,keV count rate of \src\ increased by a factor of 1.28, and the 2--4\,keV count rate increased by a factor of 1.05.

\section{Spectral Analysis}
\label{analysis}

In this section, we present an analysis of the X-ray spectra in both the high and low flux states. To study how the spectral components are changing along with the X-ray flux, we extract spectra from four intervals as indicated in panel (3) of Fig.~\ref{lcurve}. The four spectra correspond to the interval before the flux drop (Spec 1), two alternating high flux periods during the long dip (Spec 2), the low flux intervals before (Spec 3) and after (Spec 4) the second short high flux state. 

The four spectra are shown in the left panel of Fig.~\ref{spec}. We can see that all spectra are soft with a prominent blackbody-like component in the soft band. There is no significant difference between Spec 1 and Spec 2. From the high flux (Spec1/2) to low flux (Spec 3/4) state, the spectra mainly change at the hard X-ray range (> 4~keV).

We do spectral analysis with \textsc{xspec} v12.11.1. We use element abundances of \cite{Anders1989} and cross-sections of \cite{Verner1996}. Errors in this work are quoted at 90\% confidence level unless specifically noted.

\subsection{Spectral models}

\begin{figure*}
    \centering
    \includegraphics[width=0.49\linewidth]{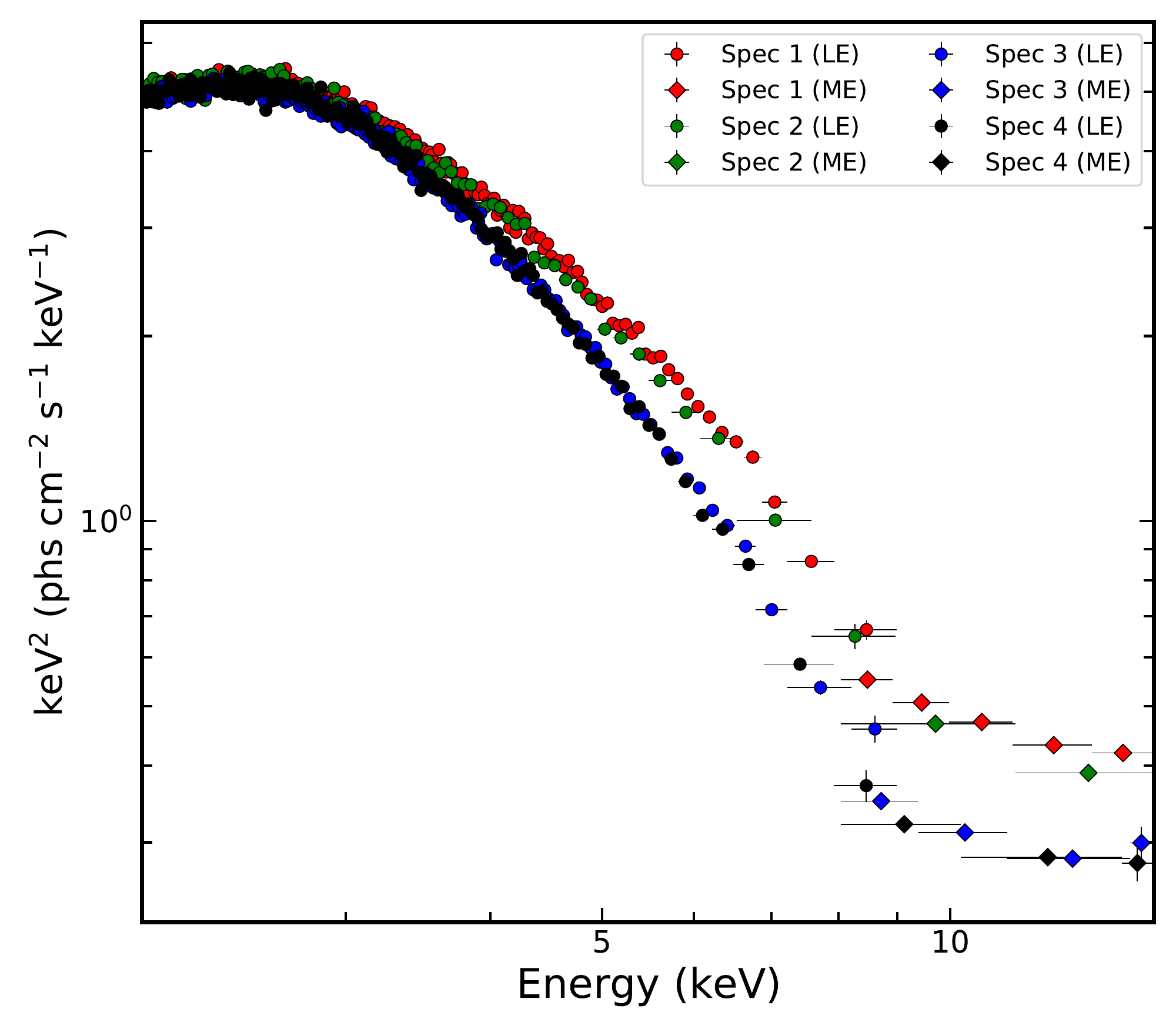}
    \includegraphics[width=0.49\linewidth]{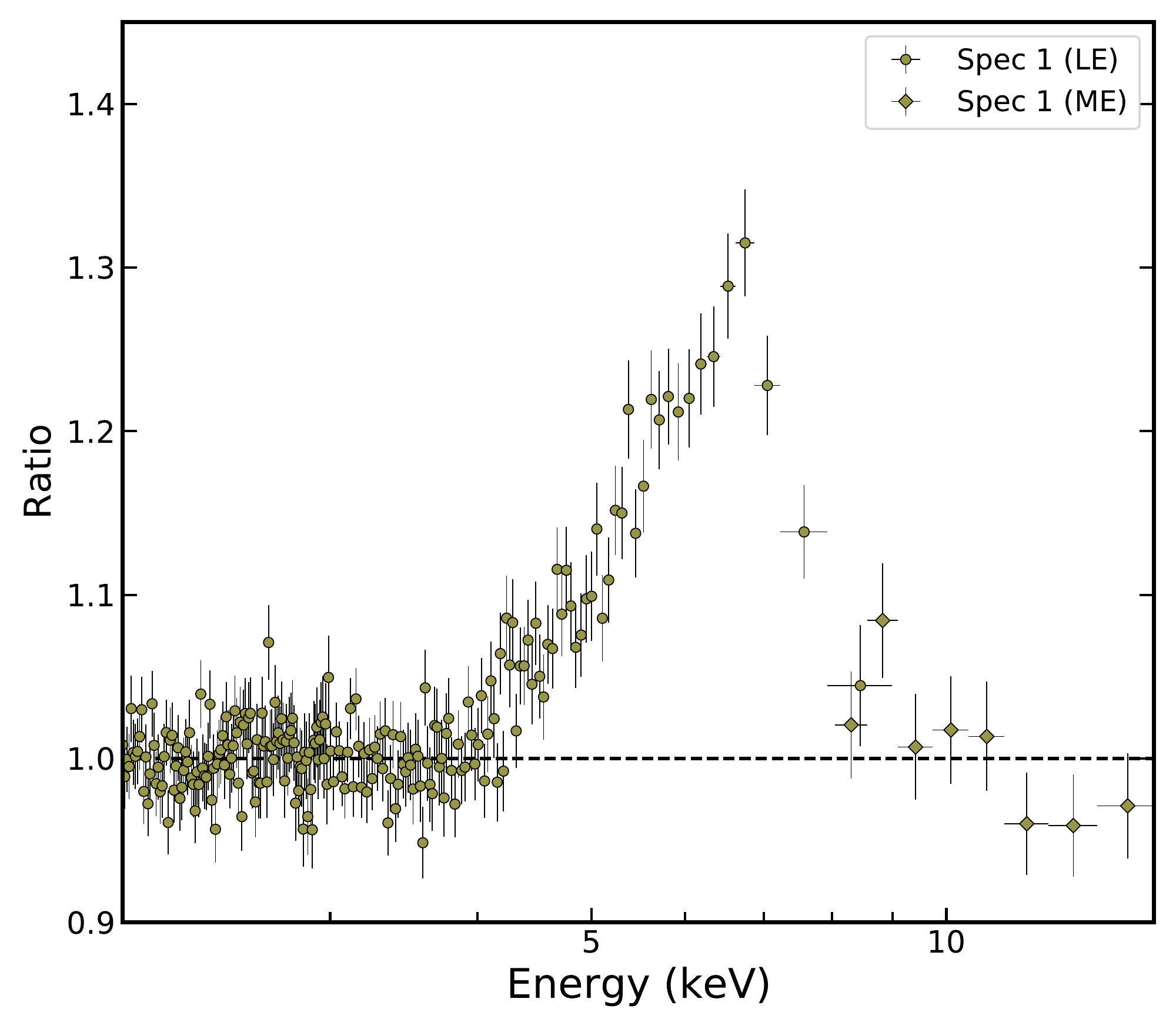}
    \caption{(Left): Spectra of the four intervals as defined in Fig~\ref{lcurve}. The LE (2--9 keV) and ME (8--15 keV) data are marked with filled circle and diamond respectively. The spectra are obtained using \textsc{setplot area} command in \textsc{xspec} and are only shown for demonstration purposes. (Right): Data to model ratio of Spec 1 to a simple absorbed continuum model. Data are binned for visual clarity.}
    \label{spec}
\end{figure*}

We first focus on Spec 1 to work out a baseline model that can be applied to all spectra. The spectra of LMXRBs often have contributions from the thermal emission from the accretion disk and power-law component from a hot corona. Thus, as the first step, we fit Spec 1 with the model \textsc{tbabs*(diskbb+nthcomp)}. The model \textsc{tbabs} \citep{Wilms2000} accounts for the Galactic absorption and has only one free parameter ($N_{\rm H}$). \textsc{diskbb} \citep{Mitsuda1984} models the thermal emission from a multi-color accretion disk and \textsc{nthcomp} \citep{Zdziarski1996, Zycki1999} fits the power-law component. Since the reflection component from an optically thick accretion disk often provides a strong contribution to the spectrum in the energy range of the iron K emission, we ignore the data in the 4--8 keV band when fitting the continuum model. The residuals of the data to the best-fit continuum model are shown in the right panel of Fig.\ref{spec}. There is evidence of a strong and broad Fe K emission line at 6--7 keV, suggesting the presence of a relativistic reflection component from the accretion disk. 

Therefore, we include a reflection component to the model and fit the four spectra simultaneously. Since the electron density of the disk of \src{} may not be constant during the outburst \citep[e.g.][]{Jiang2019b}, we implement the reflection model \textsc{reflionx} \citep{ross2007} that has a variable electron density. \textsc{reflionx} describes only non-relativistic reflection from the accretion disk, we still need to include the broadening kernel \textsc{relconv} \citep{Dauser2010, Dauser2013}, to account for the relativistic effects. The full model is: \textsc{tbabs*(diskbb+nthcomp+relconv*reflionx)} in XSPEC notation.

The spin of the black hole is fixed at the maximum value to estimate the inner radius of the disk using the reflection model. The assumption of a high BH spin is consistent with previous measurements \citep[e.g.][]{Parker2016}.
The iron abundance is fixed at the solar value in this model. The electron temperature ($kT_{\rm e}$) of the power-law component is fixed at 100~keV since we cannot constrain this parameter. We assume a power-law emissivity profile and let free the index ($q$). This model provides a good fit for Spec 1 ($\chi^2/\nu$=846/884). We have tested a broken power-law emissivity profile by fixing the outer index at 3 and letting the breaking radius ($R_{\rm br}$) free to vary. This emissivity profile improves the $\chi^2$ by only 2 and there is no significant impact on the best-fit of other parameters. Therefore, we conclude that a power-law emissivity profile is enough to fit the data. We note that there is a narrow peak in the iron line profile (see the right panel of Fig.~\ref{spec}). The narrow core is at 6.7~keV, which excludes the origin of reprocessing by distant neutral reflector. This is supported by the fact that adding a non-relativistic reflection component does not improve the fit. The line energy is consistent with the emission of Fe~XXV. It may originate from highly ionized material in the system, i.e., the accretion disk wind outside the line of sight. It is not possible to test its origin with current data. Moreover, the narrow line should not affect the analysis of the relativistically broadened iron line profile.

We obtain an inner disk temperature of 0.73~keV of the \textsc{diskbb} component that dominates the soft X-ray band (60\% of flux in 2--15 keV). The fit also needs a very soft ($\Gamma>2.9$) power-law component. From the relativistic reflection model, we measure the inclination angle of the inner disk to be $\sim$ 40 degrees. Moreover, the inner edge of the disk is slightly truncated at 2~$R_{\rm g}$ ($R_{\rm g}=GM/c^2$ is the gravitational radius) assuming a maximum black hole spin ($a_*=0.998$).

We find that the index of the emissivity profile is steeper than the canonical value of the Newtonian limit ($q=3$). This could be a result of light bending when the corona is close to the black hole. We thus replace the kernel \textsc{relconv} with \textsc{relconv\_lp}. In this model, the corona is assumed to be a point source located at a certain height ($h$) above the black hole and the emissivity profile is self-consistently calculated. This model returns a fit statistic of $\chi^2/\nu$=844/884. We indeed find that the corona height is within 3~$R_{\rm g}$ to the black hole at 90\% confidence level.

We will further discuss the geometry of the innermost accretion region, including $R_{\rm in}$ and $h$, in the following section.

\subsection{Spectral variability}

So far, we have obtained a good model for Spec 1 which includes the thermal emission and the reflection from the accretion disk. We apply the same model to Spec 2--4 to study the spectral variability of \src{}. The inclination of the disk and the column density of the Galactic absorption are not expected to change on this short timescale, so we link these parameters among different spectra. The electron density and the inner radius of the disk are also linked.

\begin{figure*}
    \centering
    \includegraphics[width=0.49\linewidth]{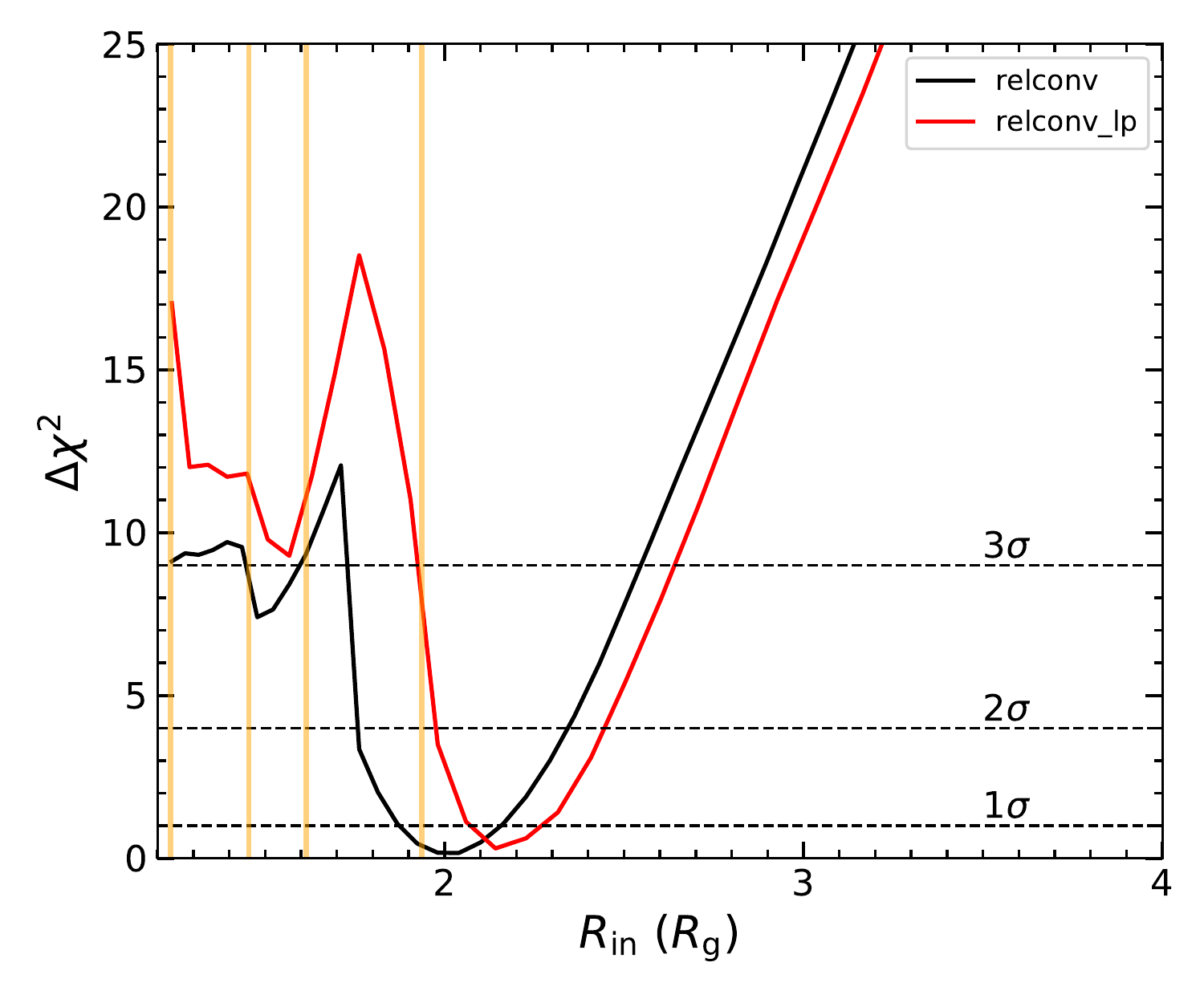}
    \includegraphics[width=0.49\linewidth]{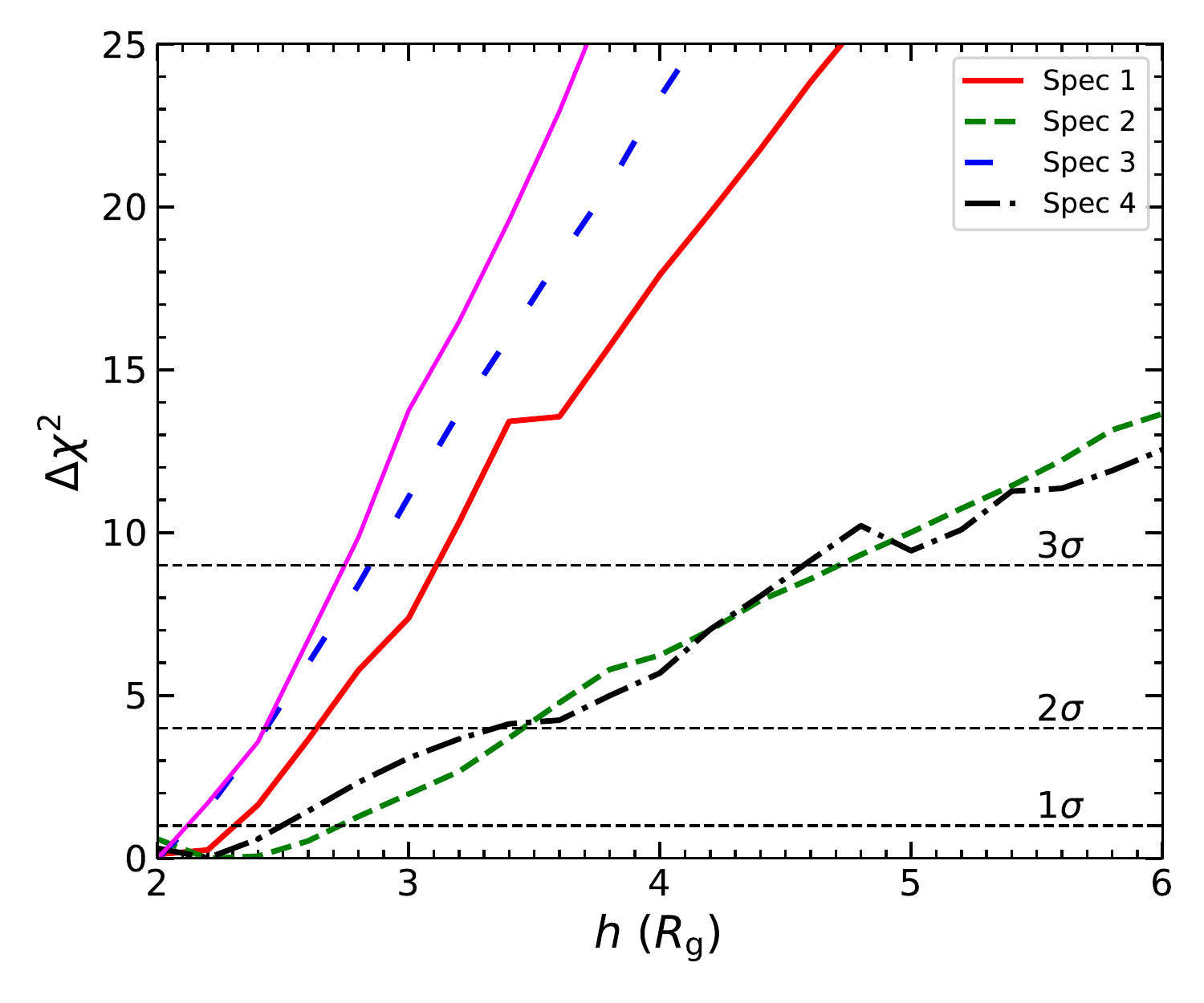}
    \caption{Left panel: The $\Delta\chi^2$ confidence contours of the inner disk radius for the two models. The orange vertical lines mark the radius of ISCO for $a_{*}=$ 0.998, 0.99, 0.98 and 0.95 respectively (from left to right). Right panel: The $\Delta\chi^2$ confidence contours for the corona height ($h$) of the four spectra. In both panels, the horizontal dashed lines represent the 1, 2 and 3$\sigma$ confidence levels. Colors are coded as in Fig~\ref{ratio}, while the magenta line represents the result when linking $h$ between the four spectra.}
    \label{delchi}
\end{figure*}

The best-fit parameters are shown in Tab.~\ref{best-fit}. The temperature of the inner disk is around 0.7~keV for both the low and high flux states. The inclination angle of the inner disk is around 40 degrees, which is consistent with previous studies \citep[e.g.][]{Garcia2015, Ludlam2015}. The electron density is only loosely constrained ($\log(n_{\rm e})<20$) but is consistent with previous studies in the very high state \citep[e.g., $18.93_{-0.16}^{+0.12}$ in][]{Jiang2019b}. The weak constraint on the density parameter is due to the limited energy range of the \hxmt\ LE data. The impact of the density parameter is confined in the soft X-ray band \citep[e.g. <3\,keV][]{ross2007,Garcia2016}. The inner disk is slightly truncated at around 2~$R_{\rm g}$. In Fig.~\ref{delchi}, we show the $\Delta\chi^2$ contour of the inner disk radius along with the location of the innermost stable circular orbit (ISCO) for some spin values. This measurement of the inner radius is consistent with ISCO ($3-\sigma$) if the spin of the black hole is 0.95.

With the \textsc{reconv\_lp} model, we find that the corona heights estimated from the four spectra are consistent within uncertainties. We have also tested the scenario with a constant corona height. The constraint on $h$ after linking it among the four spectra is shown by the black line in the right of Fig.~\ref{delchi}. We can see that the corona is required to be close to the black hole ($<3~R_{\rm g}$). 

\begin{figure*}
    \centering
    \includegraphics[width=0.49\linewidth]{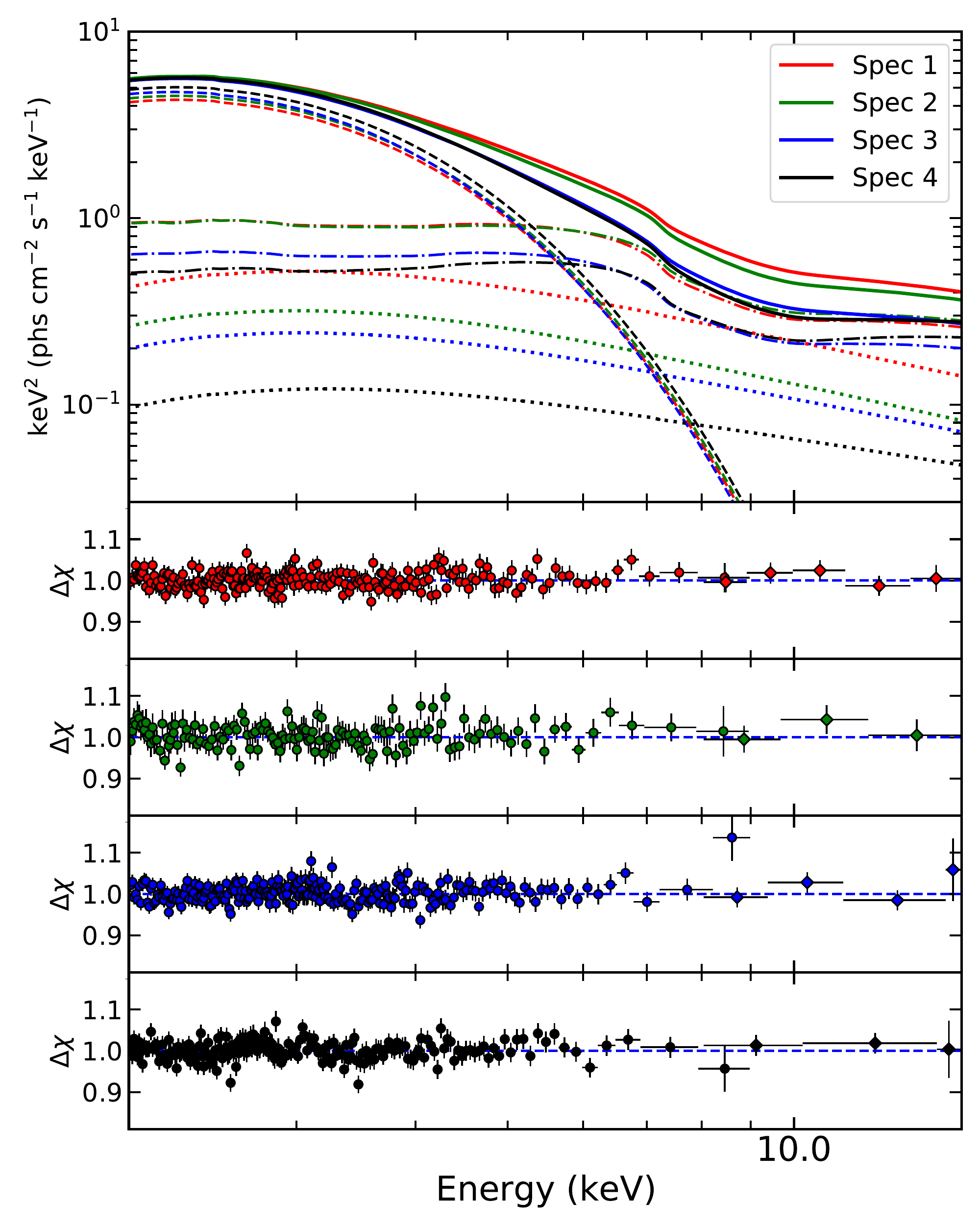}
    \includegraphics[width=0.49\linewidth]{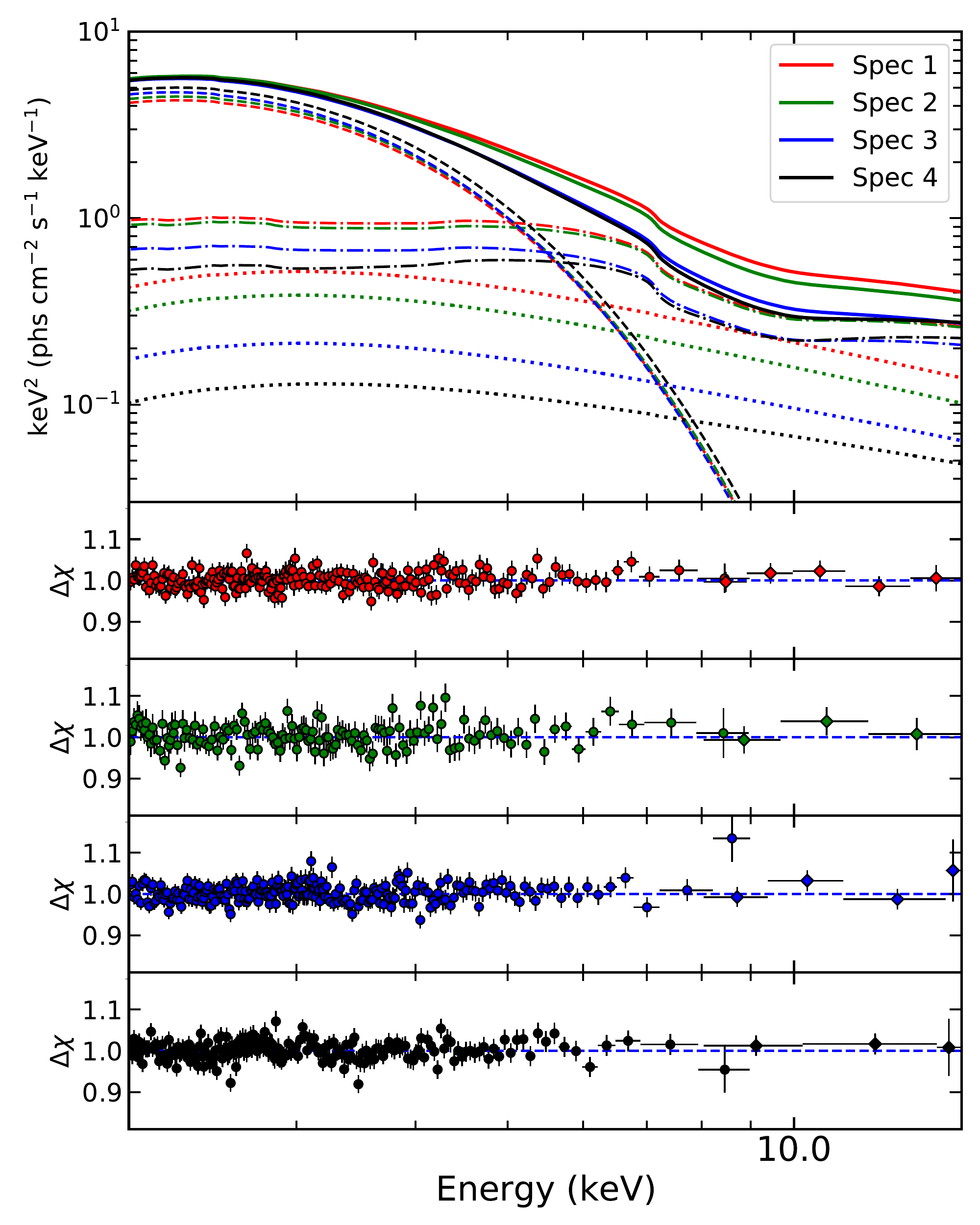}
    \caption{(Left): The best-fit models and the corresponding residuals to the four spectra with model \textsc{tbabs*(diskbb+nthcomp+relconv*reflionx)}. The LE and ME data are marked with filled circle and diamond markers respectively. (Right): The same as the left plot but with the relativistic broadening kernel replaced by \textsc{relconv\_lp}. The full models are plotted in solid lines. The disk, power-law and reflection components are plotted in dashed, dotted and dash--dotted lines respectively.}
    \label{ratio}
\end{figure*}

\begin{figure*}
    \centering
    \includegraphics[width=\linewidth]{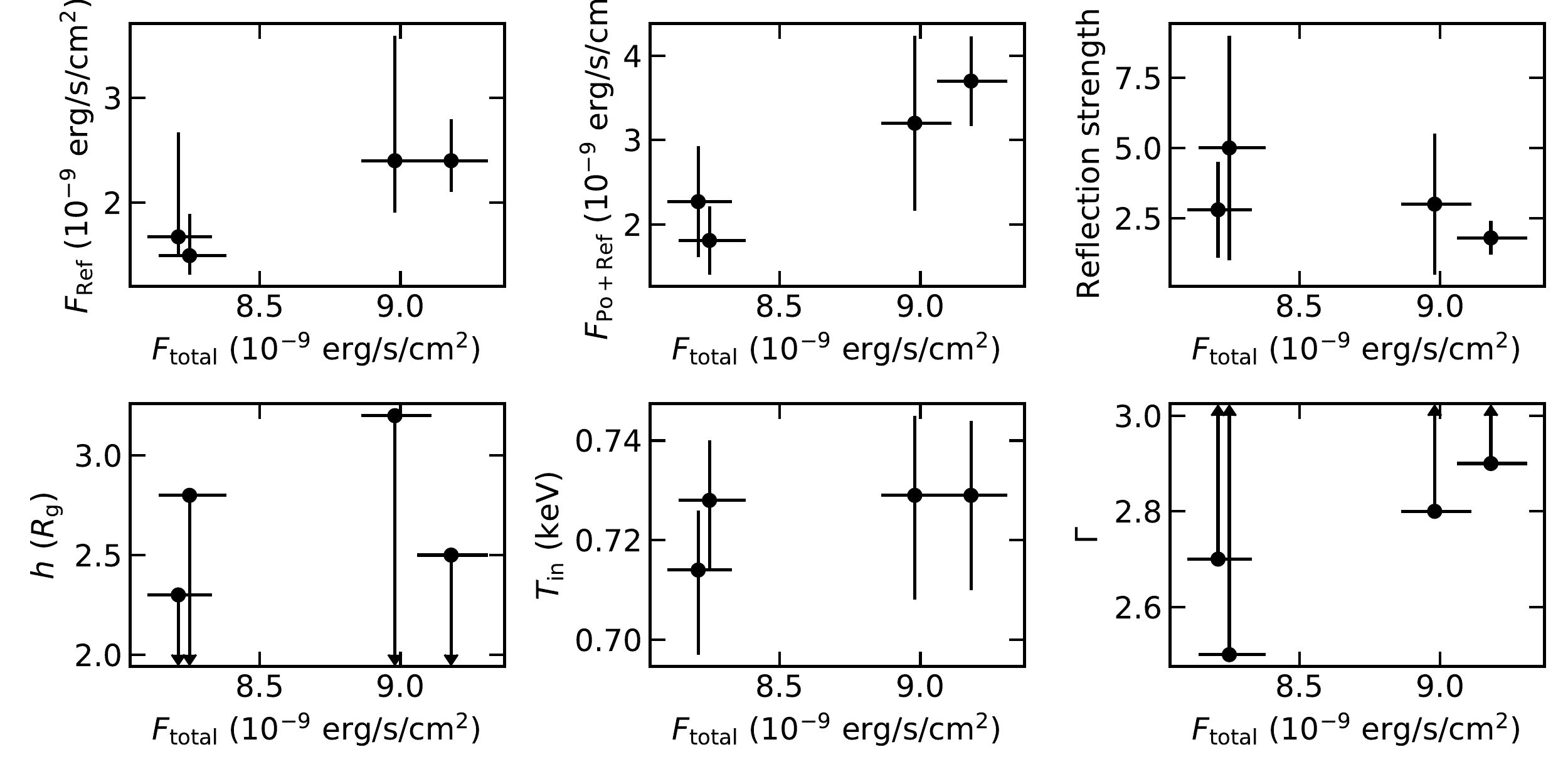}
    \caption{Relations of the some spectral parameters with the total X-ray flux in the energy range 2--15~keV. Lower and upper limits are marked with arrows.}
    \label{relations}
\end{figure*}

The best-fit model and corresponding spectral components are shown in Fig.~\ref{ratio}, along with the residuals. We can see that the changes in the \textsc{diskbb} component are almost negligible. Prominent changes are on the reflection and power-law components, which account for the difference in the hard X-ray band among the four spectra. We show the evolution of some spectral parameters with the X-ray flux of the source in Fig.~\ref{relations}. The data require a very soft ($\Gamma>2.5$) index for the power-law component but it is impossible to see how $\Gamma$ changes because of the large uncertainty. The flux of the power-law emission decreases by a factor of 4 from the high flux to the low flux states. The flux of the reflection component also decreases, but only by a factor of 1.6. The reflection strength, which is defined as the ratio between the reflection and power-law components, shows tentative evidence to increase from the high to low flux states, but the evolution is not significant if considering the uncertainty.

\begin{table*}
    \centering
    \caption{Best-fit values of the four spectra.} \label{best-fit}
    \renewcommand\arraystretch{1.5}
    \resizebox{18cm}{!}{
    \begin{tabular}{lccccc|cccc}
    \hline\hline
    & & \multicolumn{4}{c|}{\textsc{relconv}} & \multicolumn{4}{c}{\textsc{relconv\_lp}} \\
    \hline
    Model & Parameter & Spec 1 & Spec 2 & Spec3 & Spec 4 & Spec1 & Spec2 & Spec3 & Spec 4 \\
    \hline
    \textsc{tbabs} & $N_{\rm H}$ (10$^{22}$~cm$^{-2}$) & \multicolumn{4}{c|}{$0.66_{-0.07}^{+0.08}$}  & \multicolumn{4}{c}{$0.68_{-0.07}^{+0.08}$} \\
    \hline
    \textsc{diskbb} & $T_{\rm in}$ (keV) & $0.729_{-0.019}^{+0.015}$ & $0.729_{-0.021}^{+0.016}$ & $0.714_{-0.017}^{+0.012}$ & $0.728_{-0.014}^{+0.012}$ & $0.724_{-0.018}^{+0.014}$ & $0.722_{-0.021}^{+0.016}$ & $0.711_{-0.015}^{+0.013}$ & $0.725_{-0.014}^{+0.013}$ \\
                   & Norm & $3000_{-300}^{+350}$ & $3200_{-300}^{+400}$ & $3700_{-300}^{+400}$ & $3600_{-300}^{+350}$ & $3100_{-300}^{+320}$ & $3300_{-350}^{+350}$ & $3800_{-330}^{+400}$ & $3650_{-300}^{+360}$  \\
    \hline
    \textsc{relconv(lp)} & $q$ & $4.6_{-0.6}^{+0.9}$ & $4.3_{-0.7}^{+1.0}$ & $4.7_{-0.7}^{+0.9}$ & $4.3_{-0.9}^{+0.9}$ & - & - & - & - \\
                    & $h$ ($R_{\rm g}$) & - & - & - & - & $<2.5$ & $<3.2$ & $<2.3$ & $<2.8$\\
                    & $a_*$ & \multicolumn{4}{c|}{$0.998^{*}$} & \multicolumn{4}{c}{$0.998^{*}$} \\
                    & $i$ (deg) & \multicolumn{4}{c|}{$41_{-5}^{+3}$} & \multicolumn{4}{c}{$37_{-3}^{+3}$} \\
                    & $R_{\rm in}$ ($R_{\rm g}$) & \multicolumn{4}{c|}{$2.01_{-0.18}^{+0.26}$} & \multicolumn{4}{c}{$2.18_{-0.18}^{+0.22}$} \\
    \hline
    \textsc{reflionx} & $\log(n_{\rm e})$ & \multicolumn{4}{c|}{$17.3_{-1.3}^{+2.7}$} & \multicolumn{4}{c}{$<20$}\\
                    & $\xi$ & $480_{-300}^{+370}$ & $230_{-80}^{+1300}$ & $480_{-300}^{+480}$ & $480_{-350}^{+2500}$ & $480_{-300}^{+500}$ & $330_{-150}^{+1100}$ & $530_{-300}^{+2100}$ & $490_{-400}^{+1300}$\\
                     & $\Gamma$ & $>2.9$ & $>2.8$ & $>2.7$ & $>2.5$ & $>2.9$ & $>2.8$ & $>2.7$ & $>2.5$\\
                     & Norm & $63_{-40}^{+290}$ & $100_{-90}^{+100}$ & $44_{-30}^{+40}$ & $36_{-24}^{+160}$ & $65_{-40}^{+220}$ & $76_{-60}^{+78}$ & $41_{-22}^{+80}$ & $37_{-21}^{+100}$\\
    \hline
    \textsc{nthcomp} & Norm & $0.38_{-0.13}^{+0.13}$ & $0.24_{-0.2}^{+0.21}$ & $0.18_{-0.11}^{+0.09}$ & $0.08_{-0.08}^{+0.08}$ & $0.39_{-0.14}^{+0.12}$ & $0.29_{-0.19}^{+0.17}$ & $0.2_{-0.1}^{+0.1}$ & $<0.17$ \\           
    \hline
    & $F_{\rm ref}$ & $2.4_{-0.3}^{+0.4}$ & $2.4_{-0.5}^{+1.2}$  & $1.67_{-0.18}^{+1.0}$ & $1.49_{-0.18}^{+0.4}$ & $2.47_{-0.11}^{+0.6}$ & $2.4_{-0.3}^{+0.5}$ & $1.78_{-0.23}^{+0.5}$ & $1.53_{-0.14}^{+0.4}$ \\
    & $F_{\rm power-law}$ & $1.3_{-0.4}^{+0.4}$ & $0.8_{-0.6}^{+0.6}$ & $0.6_{-0.3}^{+0.27}$ & $0.32_{-0.28}^{+0.28}$ & $1.3_{-0.4}^{+0.3}$ & $0.9_{-0.7}^{+0.5}$ & $0.53_{-0.5}^{+0.14}$ & $0.34_{-P}^{+0.26}$\\
    & $F_{\rm total}$ & $9.18_{-0.12}^{+0.13}$ & $8.98_{-0.12}^{+0.13}$ & $8.21_{-0.11}^{+0.12}$ & $8.25_{-0.11}^{+0.13}$ & $9.21_{-0.12}^{+0.11}$ & $9.00_{-0.15}^{+0.11}$ & $8.24_{-0.11}^{+0.12}$ & $8.27_{-0.12}^{+0.12}$ \\
    \hline
    Fraction & ref/power-law & $1.8\pm 0.6$ & $3.0\pm 2.5$ & $2.8\pm 1.7$ & $5.0\pm 4$ & $1.9\pm 0.6$ & $2.7\pm 1.8$ & $3.4\pm 2.1$ & $4.5\pm 4$ \\
    Fraction & (ref+power-law)/total & $0.40\pm 0.06$ & $0.36\pm 0.11$ & $0.28\pm 0.08$ & $0.22\pm 0.05$ & $0.41\pm 0.05$ & $0.37\pm 0.08$ & $0.28\pm 0.05$ & $0.23\pm 0.05$ \\
    \hline
    $\chi^2$/d.o.f &  & \multicolumn{4}{c|}{3138/3380} & \multicolumn{4}{c}{3134/3380} \\
    \hline\hline 
    \end{tabular}
    }
    \textit{Note.} The flux (2--15 keV) of the reflection component and the full model are presented in units of 10$^{-9}$ erg~s$^{-1}$~cm$^{-2}$. 
\end{table*}

\section{Discussion}
\label{discussion}

At the end of the hard to soft transition of the 2021 outburst of \src{}, \hxmt{} captured two brief high flux states (Fig.~\ref{lcurve}) which lasted for a period comparable to the orbital period of the telescope. This suggests that the source was rapidly alternating between two flux states. The change of flux is more prominent in the hard X-ray band (Fig.~\ref{spec}), which excludes variable local absorption as the origin. 

To find out what emission components are changing along with the source flux, we divided the two-day observation into four intervals based on alternating properties and decomposed the spectral components for each interval. To fit the broadband spectra, we need to include disk thermal emission, power-law emission from the hot corona, and relativistic reflection component by the optically thick accretion disk. Our best-fit models suggest that it is the corona emission and the disk reflection component that drive the flux change in the hard X-ray band (Fig.~\ref{ratio}). The disk thermal emission is, however, relatively constant. 

\subsection{The inner radius of the disk}

By fitting the relativistic reflection model to the data, we find that the disk may be slightly truncated at 2~$R_{\rm g}$ when assuming a maximum black hole spin ($a_*=0.998$)\footnote{If the spin is 0.95, which is in agreement with the previous measurement with relativistic reflection \citep{Parker2016, Jiang2019b}, the disk is consistent with not being truncated.}. The inner disk radius estimated from the normalization parameter of \textsc{diskbb} is around $4\times(D/8~{\rm kpc})(10M_{\odot}/M_{\rm BH})~R_{\rm g}$ assuming a color correction factor of 1.7 and inclination of 40 degrees. Implementing the median value of the measurement of \cite{Zdziarski2019} ($M=7.5M_{\odot}$, $D=10~{\rm kpc}$) returns a inner radius of 6.6~$R_{\rm g}$. This value is larger than that obtained from modeling the relativistic reflection features. We note that there is still large uncertainty on current measurements of the distance and black hole mass of the system. If we consider the lower limit of the distance derived by \cite{Heida2017} (5~kpc), we obtain an inner radius size of 2.5~$R_{\rm g}$, which is close to the measurement from relativistic reflection.

There is a tentative trend that the disk inner radius measured with the disk component is larger in the low flux state, but the uncertainties are too large to obtain a robust conclusion. The change of the disk flux in the 2--15 keV band is less than 15\% from Spec 1 to Spec 4. The temperature of the inner edge of the disk is even more stable. Similarly, the inner radius of the disk measured by the reflection model is constant during the period of our observations. During the rapidly alternating flux states, the inner disk radius remains constant within our measurement uncertainty.

\subsection{The geometry of the corona during the state transition}

In this work, we consider a power law-shaped emissivity profile and a consistent emissivity model for the lamppost geometry. The best-fit value of the power-law index ($q$) is in the range 4--5 and is consistent between the four spectra. The index is steeper than what is expected by the Newtonian limit ($q=3$). This can happen when the corona is close to the black hole and photons are concentrated near the black hole as a result of light bending. This scenario is supported by the results with \textsc{relconv\_lp}, where the upper limit of the corona height is found to be $3.2~R_{\rm g}$.

As for the corona emission, its flux decreases by a factor of 4 from Spec 1 to Spec 4, along with the decrease of the reflection component by a factor of 1.6. One interesting thing to note is that the best-fit value of the reflection strength, namely the ratio between reflected and directly observed corona emission, increases from the high flux to low flux states. Still, the uncertainties are too large for conclusive determination. Given that the corona height is consistent between the four spectra, the reflection strength may be also constant. This suggests that the alternating states we are seeing are a result of the variation of the corona power.

However, it is impossible to see if the photon index changes between the low and high flux states. This is because we do not have data above 15~keV and the soft X-ray band is dominated by the disk emission.

\subsection{The power of the corona}

We analyzed the \hxmt\ data of \src\ during the intermediate state. During the period of the observations, \src\ quickly alternated between high and low flux states in the hard X-ray band while the soft X-ray emission did not vary significantly. The high flux state only lasted briefly for a period that is comparable to the orbital period of the telescope. The transition happened between orbits of \hxmt\ and was not observed by our observations. Readers may refer to Appendix A for the \swift\ observation of the transition.

The \hxmt\ observations of \src\ show evidence of both the disk reflection and the thermal emission from the inner accretion disk. Detailed analysis shows that the thermal emission of the disk does not change during the state transition as suggested by the unchanging soft X-ray emission (see Fig.~\ref{spec}). The variable non-thermal emission, including the power-law continuum from the corona and the reflected emission from the disk, is responsible for the hard X-ray flux transition.

Detailed reflection modeling suggests that the geometry of the inner accretion region does not change. For instance, the inner radius of the accretion disk remains close to ISCO at $2~R_{\rm g}$. Assuming a maximum BH spin, our results suggest that the disk might be slightly truncated. It is important to note that our measurement is still consistent with a non-truncated disk within a 3-sigma uncertainty range.

The geometry of the coronal region does not change either during the observed rapidly alternating flux states. The emissivity profiles of the disk, modeled by both a power law and a lamppost model, do not change significantly. There is tentative evidence that the reflection strength is higher in the high flux state than in the low flux state, but the difference is not significant. The measurements are consistent within the uncertainty ranges. This suggests that the geometry of the corona in \src\ remains constant.

The sudden increase and decrease of the coronal power law observed by \hxmt\ observations might be due to changes in the heating mechanism of the corona: the hard X-ray continuum of BH XRBs originates in the up-scattering Comptonisation processes of lower energy disk photons \citep[e.g.][]{maisack93}. A heating mechanism must exist to maintain the corona. One of the most accepted models is through the magnetic field. For instance, pair productions in the magnetosphere around a BH may play an important role in this process \citep{hirotani98}. The disk is relatively stable during the observed state transition, sudden changes in the magnetic field may release more accretion power from the disk to the corona.

Future X-ray observations of these rapidly alternating flux states in the intermediate transition may help to solve the mystery. It is very important to observe the exact moments of the rapid state transitions. Unfortunately, most of the current X-ray missions are not able to offer a non-stop monitoring program due to their low-Earth orbits. Future X-ray missions, e.g. \textit{Arcus} \citep{smith20} on a high Earth orbit, will be able to continuously observe \src\ for several days.

\subsection{The reflection component}

The strong relativistic reflection features enable us to constrain important parameters like the radius of the disk inner edge and the corona height. Another parameter that is of importance is the electron density of the accretion disk since it may affect the fit in the soft X-ray band \citep{ross2007}. The constraint on the electron density obtained with the \textsc{reflionx} model is $\log(n_{\rm e})<20$. We note that the model \textsc{relxillD} \citep{Garcia2016} also allows the electron density parameter to vary between $15<\log(n_{\rm e})<19$. This model treats the atomic physics more accurately than \textsc{reflionx}. Therefore, we test if the choice of reflection models would affect our results by applying the model \textsc{relxilllpD} to the four spectra simultaneously. The best-fit parameters are listed in Tab.~\ref{relxilllpd}.

We can see that, compared to the fit with \textsc{reflionx}, there are no significant changes on the measurements of the disk inner radius, the corona height or the inclination angle. As for the electron density, we can only get a upper limit of $\log(n_{\rm e})=16.7$. This measurement is generally consistent with that from the \textsc{reflionx} model (see Tab.~\ref{best-fit}). We can also calculate the electron density from the definition of the ionization parameter ($\xi=4\pi F_{\rm irr}/n$, where $F_{\rm irr}$ is the irradiating flux on the disk and $n$ is the electron density). Following equation (8) in \cite{Zdziarski2020}, the electron density we obtain is as high as $\sim 10^{22}$~cm$^{-3}$ for Spec1 (assuming the reflection fraction ${\cal{R}}=1$). This value is several orders of magnitude higher than that from the spectral fitting. The large discrepancy can not be explained by the uncertainties in the irradiating flux or other parameters. Further analysis of reflection spectra in the soft state are needed to explain the disagreement between the two methods.

\subsection{Compare with other sources}

The fast alternating flux states we found is similar to the ``flip-flop'' transitions. Flip-flops were first detected in \src{} \citep{Miyamoto1991} where they appear as abrupt transitions between a higher and a lower flux levels on a timescale of minutes. The phenomenon is not common and have only been found in a few other sources \citep[see][and references there in]{Bogensberger2020}. Although the light curve of \src{} we are seeing resembles the flip-flops, we note that our results are different from what have been found about flip-flops. Flip-flops are usually found during or after the very high state \citep[e.g.][]{Takizawa1997, Casella2004} while the transitions reported in this work are before the soft state. This might explain why we are seeing strong reflection features while they are not commonly seen in previous studies of flip-flops. Changes in the power spectra and quasi-periodic oscillations (QPOs) are often associated with the flip-flop transitions \citep[e.g.][]{Nespoli2003, Bogensberger2020}. However, we do not see any QPOs and the power spectra look like that of a typical soft state which almost have no power. In this sense, the transitions we are seeing are similar to the ``late flip-flops'' found in Swift~J1658.2--4242 \citep{Bogensberger2020}.

Spectral fitting of the flip-flops in Swift~J1658.2--4242 has shown that the change of the inner disk temperature is responsible for the spectral variability. On the contrary, we find a nearly consistent value of the temperature for both the low and high flux states. It is the corona power that is changing when the disk remains stable. Similar behaviors have been found during the soft state of some X-ray binaries \citep[e.g.][]{Churazov2001, Buisson2021}.

\section{Conclusions}

During the 2021 outburst of \src{}, we found that the source rapidly alternates between a low and high flux state. Spectral analysis on its time-resolved spectra shows that:

\begin{enumerate}

\item The flux change is confined in the hard X-ray band (> 4~keV) and is regulated by the power-law emission and the reflected emission.
\item The strength of the disk thermal emission and the inner radius of the accretion disk do not change with the flux.
\item The corona is close to the black hole ($3~R_{\rm g}$). There is no evidence for changes in the corona geometry when the flux changes, which suggests the change is in the intrinsic power of the corona.
\end{enumerate}

\section*{Acknowledgements}

We thank Yanfei Jiang for insightful discussions. The work of H.L, Z.Z, and C.B. is supported by the National Natural Science Foundation of China (NSFC), Grant No. 11973019, the Natural Science Foundation of Shanghai, Grant No. 22ZR1403400, the Shanghai Municipal Education Commission, Grant No. 2019-01-07-00-07-E00035, and Fudan University, Grant No. JIH1512604. J.J. acknowledges support from the Leverhulme Trust, the Isaac Newton Trust and St Edmund's College, University of Cambridge. L.J. thanks the support from the National Natural Science Foundation of China grants No. 12173103, U2038101, U1938103, 11733009, and the Guangdong Major Project of the Basic and Applied Basic Research grant 2019B030302001. S.Z. and L.K. thank the support from the National Key R\&D Program of China (2021YFA0718500) and the National Natural Science Foundation of China, Grant No.~U1838201.

\section*{Data Availability}

All the data can be downloaded from the website of \hxmt{}: http://hxmten.ihep.ac.cn/. Please contact John Tomsick (jtomsick@berkeley.edu) for the \textsc{reflionx} model.




\bibliographystyle{mnras}
\bibliography{bibliography} 




\appendix
\section{}
\begin{figure*}
    \centering
    \includegraphics[width=\linewidth]{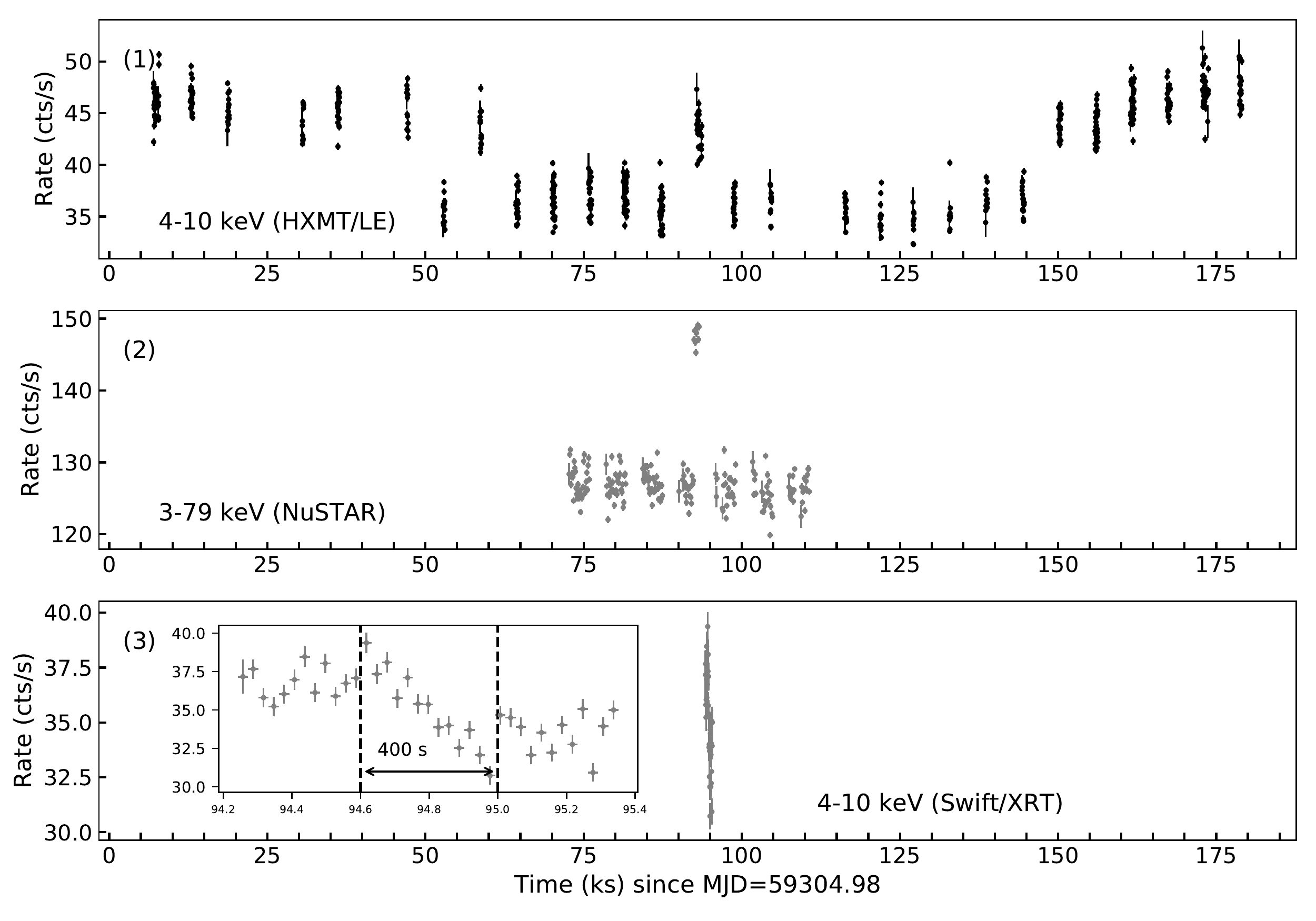}
    \caption{Light curves of \src{} by \hxmt{}, \nustar{} and \swift{}.}
    \label{otherlc}
\end{figure*}

The \nustar{} light curve is extracted from the observation on 2021 April 1st with obsID: 90702303011. The data are first processed with the tool \textsc{nupipeline} ver 0.4.8. We then extract the light curve using the tool \textsc{nuproducts} with the source region being a 180 arcsec circle. The background product is extracted from a circular region with the same size near the source.

The \swift{}/XRT light curve is extracted from the observation (obsID: 00014218002) on the same day. Data are processed with tools \textsc{xrtpipeline} ver 0.13.5 and \textsc{xselect}. Source light curve is extracted from a circular region of 100 arcsec and the background region is near the source with the same size.

As shown in Fig.~\ref{otherlc}, the brief high flux state at 93~ks is captured by \hxmt{}, \nustar{} and \swift{}. In particular, \swift/XRT captures the continuous flux drop from the high to the low flux state (panel (3) of Fig.~\ref{otherlc}), which takes around 400~s. However, the exact moment of the flux increase is not captured. 

\begin{table*}
    \centering
    \caption{Best-fit values of the four spectra with the model \textsc{relxilllpD}.} \label{relxilllpd}
    \renewcommand\arraystretch{1.5}
    \begin{tabular}{lccccc}
    \hline\hline
    \hline
    Model & Parameter & Spec 1 & Spec 2 & Spec3 & Spec 4 \\
    \hline
    \textsc{tbabs} & $N_{\rm H}$ (10$^{22}$~cm$^{-2}$) & \multicolumn{4}{c}{$0.93_{-0.06}^{+0.07}$}   \\
    \hline
    \textsc{diskbb} & $T_{\rm in}$ (keV) & $0.731_{-0.016}^{+0.016}$ & $0.728_{-0.019}^{+0.019}$ & $0.704_{-0.013}^{+0.012}$ & $0.718_{-0.014}^{+0.013}$  \\
                   & Norm & $2700_{-200}^{+260}$ & $3000_{-260}^{+350}$ & $3700_{-300}^{+360}$ & $3600_{-280}^{+350}$ \\
    \hline
    \textsc{relxilllpD} & $h$ ($R_{\rm g}$) & $<2.5$ & $<3.1$ & $<2.1$ & $<2.6$\\
                    & $a_*$ & \multicolumn{4}{c}{$0.998^{*}$}  \\
                    & $i$ (deg) & \multicolumn{4}{c}{$39.5_{-2.7}^{+2.7}$}  \\
                    & $R_{\rm in}$ ($R_{\rm g}$) & \multicolumn{4}{c}{$2.16_{-0.11}^{+0.19}$}  \\
                    & $\log(n_{\rm e})$ & \multicolumn{4}{c}{$<16.7$} \\
                    & $\log(\xi)$ & $4.22_{-0.11}^{+0.21}$ & $4.15_{-0.11}^{+0.3}$ & $4.20_{-0.26}^{+0.2}$ & $3.99_{-0.25}^{+0.13}$ \\
                    & $\Gamma$ & $>2.9$ & $>2.9$ & $>2.8$ & $>2.9$ \\
                    & Norm & $10_{-3}^{+120}$ & $6_{-3}^{+8}$ & $50_{-30}^{+70}$ & $9_{-5}^{+2}$ \\
    \hline
    \textsc{nthcomp} & Norm & $0.30_{-0.12}^{+0.12}$ & $0.18_{-0.18}^{+0.19}$ & $<0.2$ & $<0.1$ \\           
    
    \hline
    $\chi^2$/d.o.f &  & \multicolumn{4}{c}{3129/3380}  \\
    \hline\hline 
    \end{tabular}

\end{table*}


\bsp	
\label{lastpage}
\end{document}